\documentclass{ws-ijmpe}

\usepackage[super,compress]{cite}

\usepackage{url}

\usepackage{bbm}

\begin{document}




\catchline{}{}{}{}{}


\title{The pionic Drell-Yan process: a brief survey}

\author{\footnotesize Wen-Chen Chang}
\address{Institute of Physics, Academia Sinica\\
Taipei 11529, Taiwan\\
changwc@phys.sinica.edu.tw }

\author{\footnotesize Dipangkar Dutta}
\address{Department of Physics and Astronomy, Mississippi State University, \\
Mississippi State, Mississippi 39762, USA\\
d.dutta@msstate.edu }
\maketitle

\begin{history}





\end{history}

\begin{abstract}
The Drell-Yan process is an invaluable tool for probing the structure
of hadrons, and the pion-induced Drell-Yan process is unique in its
sensitivity to several subtleties of the partonic structure. This is a
brief review of the pionic Drell-Yan process with particular emphasis
on the dilepton angular distributions and nuclear modification of the
parton distributions (the EMC effect).
\end{abstract}


\ccode{PACS numbers: 12.38.Qk, 13.85.Qk, 13.88.+e, 14.20.Dh, 21.65.Cd, 24.85.+p, 25.80.Hp}










\section{Introduction}

\label{sec:intro}

The observation of a continuum of massive lepton pairs in
proton-uranium collisions~\cite{christenson70,christenson73} was explained by S. Drell
and T.-M. Yan~\cite{drell-yan1,drell-yan2} by the process that now bears their
name. A leading order Drell-Yan parton model of collinear
quark-antiquark ($q\bar{q}$) electromagnetic annihilation into a
virtual photon successfully described the scaling behavior, the
$A$-dependence of production cross section, and the polar angle
distribution of leptons. Furthermore, the inclusion of the
perturbative QCD effect of gluon emission and absorption by the
partons could explain nicely, the transverse momentum distribution of
lepton pairs and the $K$ factor (ratio of next-to-leading order to
leading order cross section) enhancement of cross section above the
naive Drell-Yan cross section. Hence Drell-Yan process has become one
of the most intensively studied processes in QCD~\cite{dyreview1,dyreview2,dyreview3} and
an effective tool to study the partonic structure of hadrons, ranging
from the Parton Distribution Functions (PDFs)~\cite{PDF} to Transverse
Momentum Dependent (TMD) distributions~\cite{TMD}. In the Drell-Yan
process the partonic structure is probed via the time-like photon
which is complementary to the space-like one in deep inelastic
scattering.

Moreover, since the production cross section of the Drell-Yan process
scales as $\tau=m_{l\bar{l}}/s=x_qx_{\bar{q}}$, where $m_{l\bar{l}}$
is the dilepton invariant mass, $s$ is the square of the
center-of-mass energy and $x_{q/\bar{q}}$ is the momentum fraction of
the hadrons carried by the struck quark/anti-quark, and, since pions
are composed of valance anti-quarks, the pion-induced Drell-Yan
process is more effective in producing lepton pairs of large
$m_{l\bar{l}}$, compared to the proton-induced process where the
anti-quarks are sea quarks abundant only at relatively small $x$. In
addition to yielding important information on the partonic structure
of the target nucleons, pion-induced Drell-Yan process plays a unique
role in exploring the pion structure, which is not accessible via deep
inelastic scattering (DIS) where a stable target is required. The
pion-induced Drell-Yan~\cite{HEPdata} reaction has been studied at
BNL, FNAL, CERN, and Dubna since 1979. In this article we review the
results from five experiments: the CIP~\cite{CIP1,CIP2,CIP3,CIP4} experiment at
BNL, the NA3~\cite{NA3,na3}, Omega~\cite{omega} and
NA10~\cite{NA10:86,NA10:88,na10} experiments at CERN and the E615~\cite{E615}
experiment at FNAL. We will emphasize their results on the angular
distributions of lepton pairs in Sec.~\ref{sec:angdist}, and the
nuclear modification of quark distributions known as the EMC effect in
Sec.~\ref{sec:emc}-\ref{sec:emcflavor}. The experimental parameters of
beam, target and range of dilepton invariant mass for these
experiments are listed in Table.~\ref{tab1}.

\begin{table}[hbt]
\tbl{Experimental parameters of pion-induced experiments}
{\begin{tabular}{@{}cccc@{}} \toprule
Experiment & Beam (Momentum GeV/$c$) & Target & $m_{l\bar{l}}$ (GeV/$c^2$) \\ \colrule
CIP & $\pi^\pm$ (80, 225, 253) & C, Cu, W & [3.0,9.0] \\
NA3 & $\pi^\pm$ (150, 200, 280) & H, Pt & [4.1,8.5] \\
NA10 & $\pi^-$ (140, 194, 286) & W, D & [4.0,8.5] \\ 
Omega & $\pi^\pm$ (39.5) & W &[2.0,2.7],[4.0,5.0] \\ 
E615 & $\pi^-$ (252) & W & [4.1,8.6] \\ 
\botrule
\end{tabular}\label{tab1}}
\end{table}

Recent experimental and theoretical studies have enabled a
multi-dimensional description of the partonic structure of nucleons:
TMD distributions and Generalized Parton Distributions~\cite{GPD}
(GPD). In Sec.~\ref{sec:future}, we will discuss possible future
measurements of the exclusive pion-induced Drell-Yan process and
measurements using a transversely polarized target. Such measurements
in parallel with DIS measurements can make unique and significant
contribution to the construction of these next-generation parton
distribution functions.


\section{Unpolarized Drell-Yan angular distribution}


\label{sec:angdist}

Despite the success of perturbative QCD in describing the Drell-Yan
cross sections, it still remains a challenge to fully comprehend the
angular distributions of dilepton pairs. Assuming dominance of the
single-photon process, the angular distribution of leptons from
unpolarized Drell-Yan process could be denoted by the angular
parameters $\lambda$, $\mu$ and $\nu$ as follows:

\begin{equation}
\frac{d \sigma}{d \Omega} \propto (1+ \lambda \cos^2\theta + \mu
\sin2\theta \cos \phi + \frac{\nu}{2}\sin^2\theta
\cos2\phi), \label{AnguDist1}
\end{equation}

where $\theta$ and $\phi$ are the polar and azimuthal angles of the
decayed leptons in the rest frame of virtual photon.

In the scheme of collinear $q\bar{q}$ electromagnetic annihilation,
the produced virtual photon is purely transversely polarized and the
angular distribution is proportional to $(1+\cos^2\theta)$,
i.e. $\lambda =1$ and $\mu=\nu=0$. The first measurement of polar
angle distributions in the t-channel helicity frame or
Gottfried-Jackson (GJ) frame by CIP experiment~\cite{CIP1} nicely
confirmed the theoretical prediction and provided a strong support of
Drell-Yan model in describing the dilepton production from the
hadron-hadron collisions.

\begin{figure}[th]
\centerline{\psfig{file=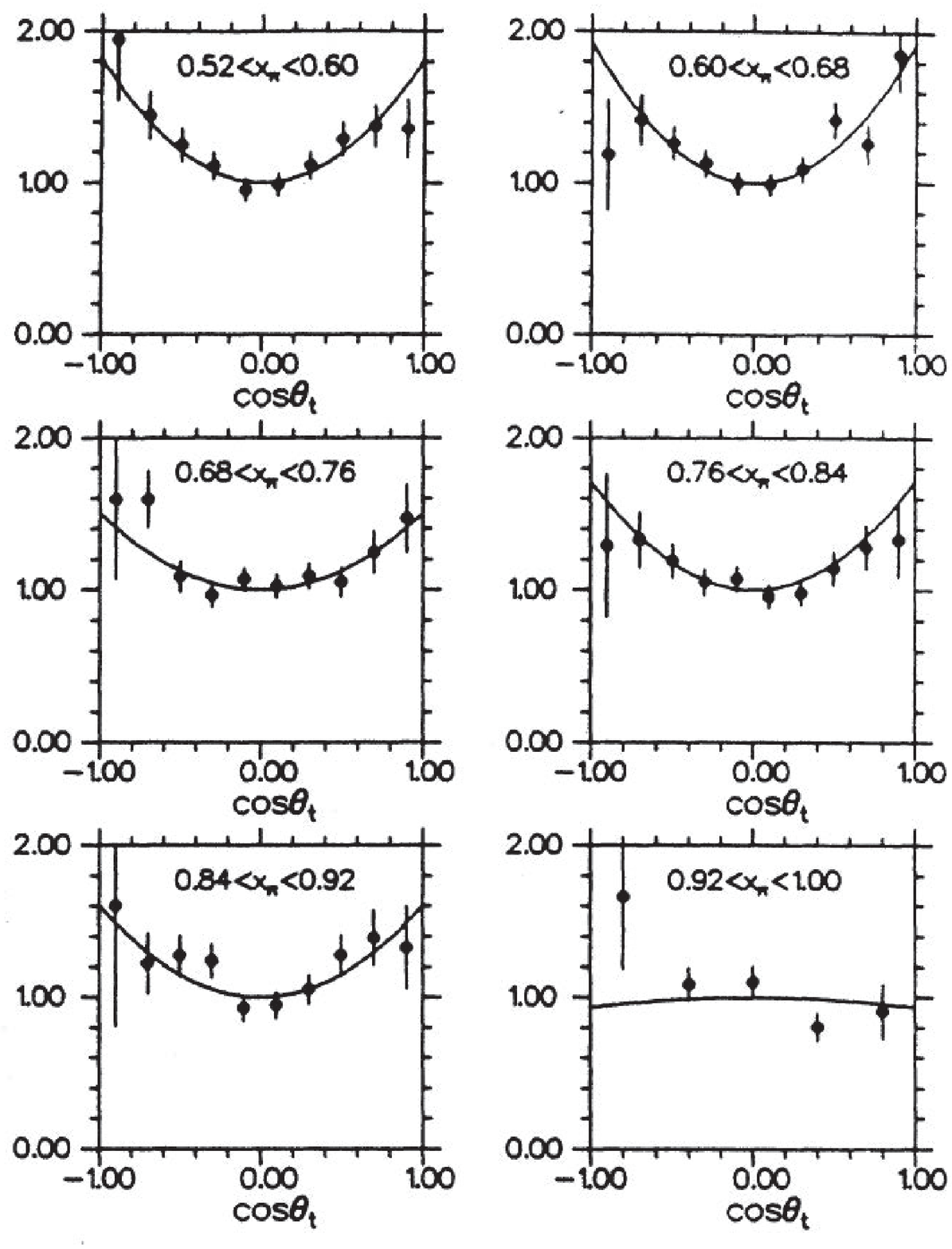,width=0.5\textwidth}\psfig{file=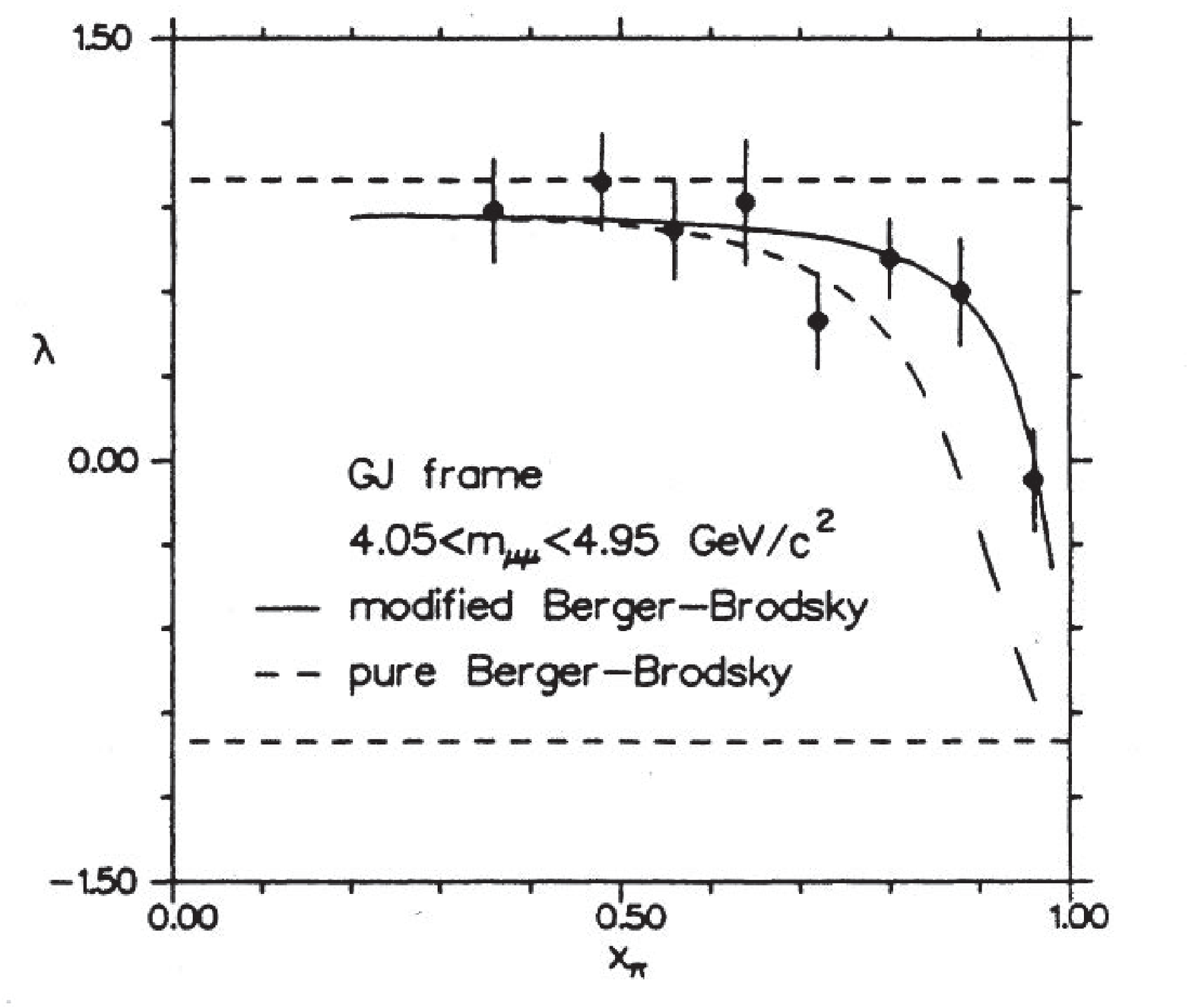,width=0.5\textwidth}}
\caption{ (a) The GJ $\cos \theta$ distribution in regions of
  $x_{\pi}$ for $4.05<m_{l\bar{l}}<4.95$ GeV/$c^2$. (b) Parameter $\lambda$
  as a function of $x_{\pi}$. The curves are from high-twist
  predictions of Ref.~\protect \refcite{Berger}. Figures taken from
  Ref.~\protect \refcite{E615}. }
\label{fig:fig_lambda}
\end{figure}

Later it was also found by the CIP~\cite{CIP2,CIP3,CIP4} experiment that the
polarization of virtual photon in GJ frame changed from transversely
polarized ($\lambda =1$) to longitudinally polarized ($\lambda=-1$)
when Bjorken-$x$ of anti-quark in the pion beam ($x_{\pi}$) approached
1. The magnitude of this variation was reduced in the Collins-Soper
(CS) frame where the effect of intrinsic quark transverse momenta is
minimized. This observation was not considered as conclusive in the
subsequent measurements performed by the NA3~\cite{NA3} and the
NA10~\cite{NA10:86,NA10:88} experiments. The controversy was finally settled by a
clear observation of this phenomenon with good statistics in the E615
experiment~\cite{E615} as shown in Fig.~\ref{fig:fig_lambda}.

In 1978, Lam and Tung studied the QCD-induced finite transverse
momentum effect on the angular distributions of lepton
pairs~\cite{LamTung78,LamTung79,LamTung80}. They found that QCD effect could lead to
$\lambda \neq 1$ and $\mu,\nu \neq 0$ but the relation
$1-\lambda=2\mu$ in CS frame, so-called ``Lam-Tung relation'' holds
for next-to-leading-order(NLO) QCD effect. Later studies showed that
the Lam-Tung relation remains almost unchanged even up to
NNLO~\cite{Nachtmann,LT_NNLO}. Therefore this relation provides a
unique opportunity to test the ``QCD-improved quark-parton
model''~\cite{LamTung78,LamTung79,LamTung80}.

\begin{figure}[hbt]
\centerline{\psfig{file=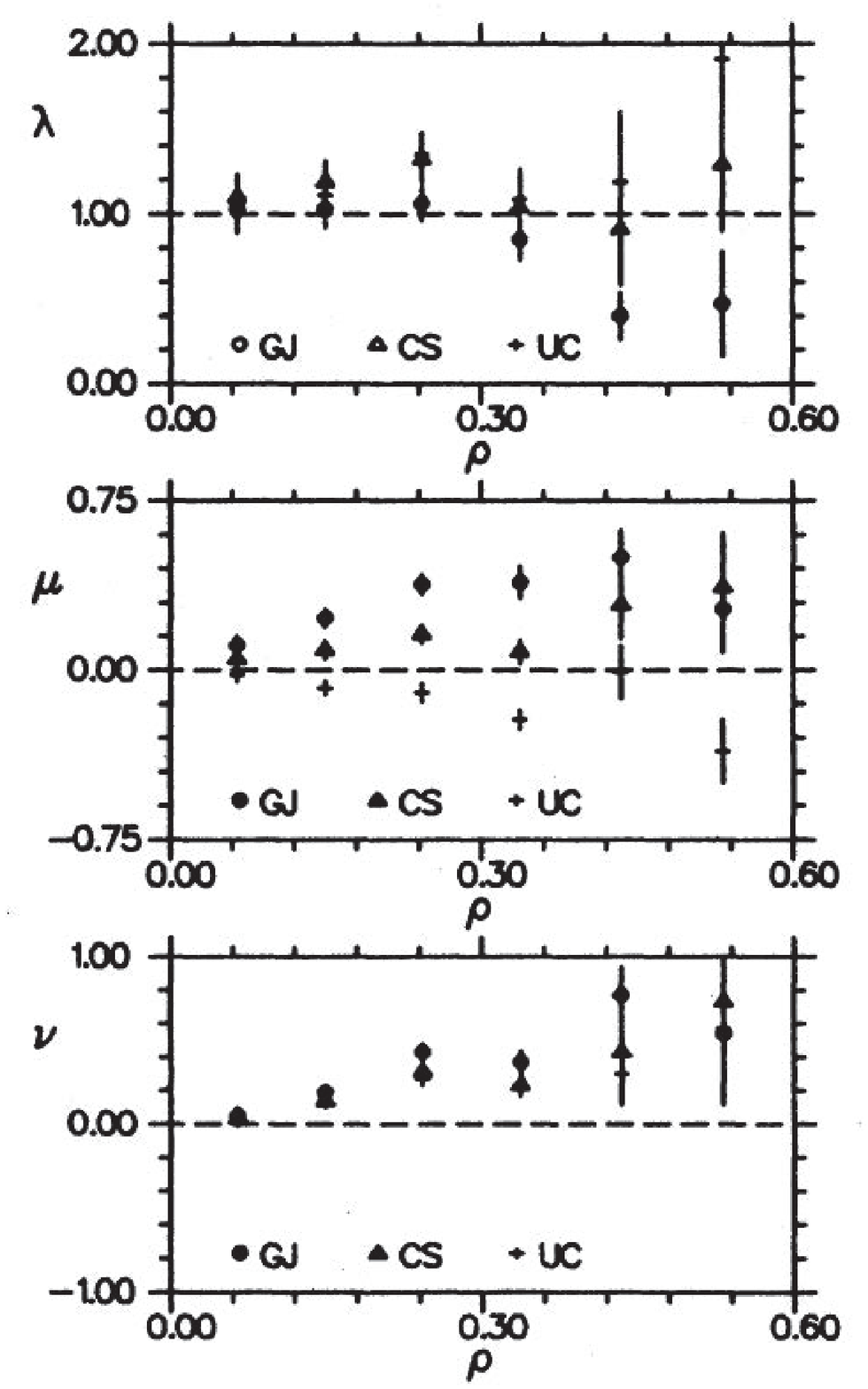,width=0.48\textwidth}\psfig{file=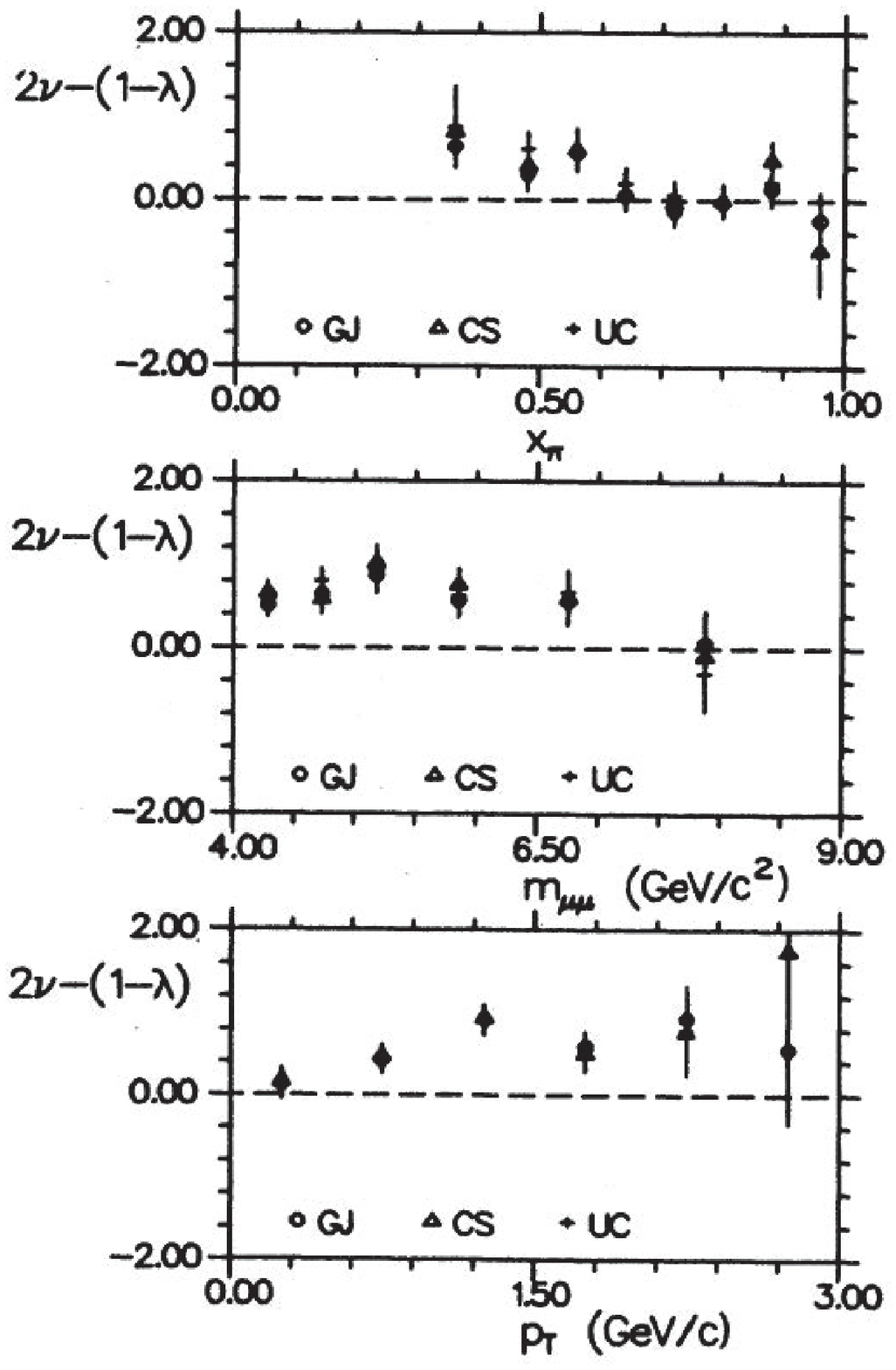,width=0.45\textwidth}}
\caption{ (a) Angular parameters $\lambda$, $\mu$ and $\nu$ as a
  function of $\rho \equiv p_T/m_{l\bar{l}}$ in GJ, CS and u-channel
  frames. (b) Test of the Lam-Tung relation $1-\lambda=2\mu$. Figures
  taken from Ref.~\protect \refcite{E615}. }
\label{fig_LTvio}
\end{figure}

The very first measurement by the NA3~\cite{NA3} experiment showed
that $\nu$ increased strongly toward large transverse momentum of the
lepton pair ($p_{T}$) but the Lam-Tung relation was roughly
preserved. Nevertheless the following measurements with better
statistics by the NA10~\cite{NA10:86,NA10:88} and the E615~\cite{E615}
experiments clearly identified a strong violation of Lam-Tung relation
in the pion-induced Drell-Yan process with nuclei and deuterium
targets. Fig.~\ref{fig_LTvio} shows E615 results of angular parameters
$\lambda$, $\mu$ and $\nu$ as a function of $p_T/m_{l\bar{l}}$ and the
degree of violation of Lam-Tung relation in the region of ~$0<p_T<5$
GeV/$c$, $4.05<m_{l\bar{l}}<8.55$ GeV/$c^2$ and $0.2<x_{\pi}<1.0$. Clearly
$\lambda$ deviates from 1 and both $\mu$ and $\nu$ have large non-zero
values. The degree of violation increases with $p_T$. The fact that
the violation was also observed using deuterium target~\cite{NA10:86,NA10:88}
excluded the possibility of a nuclear target effect.

\subsection{Theoretical Interpretations} 

\label{sec:angdist_theory}


As described above, there are two phenomena in the angular
distributions of pion-induced Drell-Yan processes which are beyond
conventional perturbative QCD: longitudinally polarized virtual
photons at large $x_{\pi}$ and violation of Lam-Tung relation at large
$p_T$. Several attempts have been made in understanding their origins
and some of them are introduced below.

\subsubsection{longitudinally polarized virtual photons at large $x_{\pi}$ }

Berger and Brodsky~\cite{Berger} explained the change of virtual
photon polarization near $x_{\pi}=1$ by higher-twist QCD effect. The
diagram of scattering of pions as two valence partons with an exchange
of gluons between each other or the pion valance anti-quark and quark
of target nucleon becomes dominant. The angular distribution was
described as follows:
\begin{equation}
\frac{d \sigma}{d \Omega} \propto (1-x_{\pi})^2(1+\cos^2\theta) +
\frac{4x_{\pi}^2 \langle k_{T}^2 \rangle
}{9m_{l\bar{l}}^2}(1-\cos^2\theta) \label{AnguDist2}
\end{equation}
where $\langle k_{T}^2 \rangle$ is the average of the square of
transverse momentum of pion valance anti-quark.

In comparison with Eq.~\eqref{AnguDist1}, $\lambda$ is $1$ at small
$x_{\pi}$ and gradually turns to be $-1$ as $x_{\pi} \rightarrow
1$. Furthermore it has been illustrated that the $x_{\pi}$-dependence
of $\lambda$, $\mu$ and $\nu$ in the angular distributions of leptons
are sensitive to the pion distribution amplitude
(DA)~\cite{Brandenburg}, which represents the distribution of
light-cone momentum fraction in the Fock state of two valence quarks
for pion. Pion DA is an important component for QCD light-cone sum
rule and QCD factorization theory in describing the exclusive
processes, e.g. $\gamma \gamma^* \rightarrow \pi^0$ the pion-photon
transition form factor at $e^+e^-$ collisions~\cite{TFF}. Recent
investigation~\cite{Bakulev} taking into account of QCD evolution of
pion DA shows that the sensitivity of dilepton angular distributions
to pion DA still remains, especially for the lepton pairs with large
$\rho$ ($\equiv p_T/m_{l\bar{l}}$).


\subsubsection{Violation of Lam-Tung relation at large $p_T$}

\label{sec:angdist2_theory}


A general two-particle spin-density matrix for the $q\bar{q}$ pair
prior to the annihilation is described by~\cite{Nachtmann,Boer05}:

\begin{equation}
\rho^{(q,\bar{q})}=\frac{1}{4}{ 
  \mathbbm{1}^{q} \otimes \mathbbm{1}^{\bar{q}} + 
  F_j \sigma_j^{q} \otimes \mathbbm{1}^{\bar{q}} + 
  G_j \mathbbm{1}^{q} \otimes \sigma_j^{\bar{q}} + 
  H_{ij} \sigma_i^{q} \otimes \sigma_j^{\bar{q}}
  } .
\end{equation}
The quantities $F_j$, $G_j$ and $H_{ij}$ characterizes different parts
of spin configurations. In the scheme of non-polarized $q$ and
$\bar{q}$, only the diagonal term survives
$\rho^{(q,\bar{q})}=\frac{1}{4}{ \mathbbm{1}^{q} \otimes
  \mathbbm{1}^{\bar{q}}}$.

The interference between opposite photon helicities gives rise to a
$\nu$ or $\cos 2\phi$ asymmetry and this means that the annihilated
$q$ and $\bar{q}$ carry specific helicity configurations. The
observation of violation of the Lam-Tung relation suggests certain
mechanisms correlating the helicities of quark and anti-quark which
comes from two individual hadrons. Such an effect is commonly expected
to be of non-perturbative QCD in origin.

Brandenburg {\it et al.}~\cite{Nachtmann} proposed a factorization
breaking effect caused by nontrivial QCD vacuum. Through the spin-flip
gluon synchrotron emission and Chromo-magnetic Sokolov-Ternov effect,
a spin correlation between an annihilating quark and anti-quark
happens~\cite{Nachtmann2,Nachtmann3}. Thus $H_{ij}$ becomes non-zero and in
general there is no factorization of the spin density matrix. It was
shown that a non-zero correlation coefficient $\kappa$ could describe
the observed Lam-Tung violation:

\begin{equation}
\kappa \equiv \frac{H_{22}-H_{11}}{1+H_{33}} \approx -\frac{1}{4}(1-\lambda-2\nu).
\end{equation}

The $q\bar{q}$ spin-density matrix becomes entangled and cannot be
factorized. Since this effect originates from the property of QCD
vacuum, it is expected to be independent of quark (anti-quark) flavor
and the momentum fraction $x$ and and the amount of violation is
predicted to persist at large $p_{T}$.

On the other hand, Boer~\cite{Boer:LT} considered a hadronic effect
where the spin direction of quark (anti-quark) correlates with its
transverse momentum ($k_{T}$) within an unpolarized hadron
itself. This correlation function ``Boer-Mulders functions''
($h_1^\bot(x,k_T^2)$) describes the unbalance of number densities of
quarks with opposite transverse polarization with respect to the
unpolarized hadron momentum~\cite{BM}. It is one of the key component
of TMD parton distributions to be extracted from unpolarized
hadrons. In this approach, the $q\bar{q}$ spin-density matrix becomes
factorized as the production of two nontrivial one-particle
spin-density matrices: $\rho^{(q,\bar{q})}=\frac{1}{4}{
  \mathbbm{1}^{q} \otimes \mathbbm{1}^{\bar{q}}}+ F_j \sigma_j^{q}
\otimes \mathbbm{1}^{\bar{q}} + G_j \mathbbm{1}^{q} \otimes
\sigma_j^{\bar{q}}$ and $\kappa = \nu/2 \propto
h_1^\bot(q_N)h_1^\bot(\bar{q}_{\pi})$. In general $h_1^\bot$ has the
quark-flavor and $x$ dependence and is parametrized to vanish at large
$k_T^2$.

Recently FNAL E866 experiment measured proton-induced Drell-Yan
process with protons~\cite{E866:p} and deuterons~\cite{E866:d} but did
not observe the violation of Lam-Tung relation as in the pion-induced
one. This result could be interpreted as the smallness of Boer-Mulders
function for sea quarks in the target nucleons but is less compatible
with the supposed flavor-blind QCD vacuum effect. In addition no
violation of Lam-Tung relation was seen in the anti-proton-induced
Drell-Yan process by CDF~\cite{CDF}. Since this measurement was made
in the region of large $m_{l\bar{l}}$ at Z-pole and very large $p_T$,
the result might not be sensitive to the proposed non-perturbative QCD
effect.

Besides the above two attempts, there are considerations of QCD
instanton-induced effect~\cite{Instanton} and Glauber gluons in the
$k_T$ factorization theorem~\cite{GlauberGluon}. As for the mechanism
of Glauber gluons, a large Glauber phase solely for the pion is
claimed and therefore it predicts that the violation of the Lam-Tung
relation would be observed only with the pion beam but not with the
other hadron beams~\cite{GlauberGluon}. More precise measurement of
dilepton angular distributions with pion beam over a wide kinematic
range and with the other hadron beams like anti-proton and kaon, will
help differentiate the theoretical interpretations.


\section{The EMC Effect}

\label{sec:emc}


In addition to probing the parton distribution functions in hadrons,
the Drell-Yan process is also sensitive to modifications to the parton
distributions for nucleons bound inside nuclei. However, since the
energy scale probed in experiments studying the partonic structure is
orders of magnitude larger than the nuclear binding energy, the
nuclear parton distributions were expected to be the same as those of
the nucleons.

In 1983, a muon-induced a Deep Inelastic Scattering (DIS) experiment
by the European Muon Collaboration (EMC)~\cite{emc1} discovered that
the nucleon structure functions measured on iron and deuterium were
different. They observed a depletion of the iron structure functions
at large $x$, which was dubbed as the ``EMC effect''. The EMC effect
has since been confirmed over a broad range of nuclear masses ($A$)
and momentum transfers ($Q^2$), by several DIS experiments using
electron~\cite{slac}, muon~\cite{emc1,muons,muons87} and
neutrino~\cite{neutrinos,neutrinos2} beams. The EMC effect is considered to be a
clear evidence that the quark distributions in nuclei are modified
compared to those in the nucleons. As mentioned, because of the high
energy scale probed in DIS experiments, the EMC effect was an
unexpected observation.  With the ever improving precision of the DIS
experiments, specially the most recent experiments on light
nuclei~\cite{hermes,seely09}, they provide a detailed and
multi-dimensional picture of the nuclear modification of the quark
distributions. It is naturally expected that the changes in quark
distributions inside nuclei should also manifest itself in other
processes, where the quark distributions are important, such as the
Drell-Yan process~\cite{drell-yan1,drell-yan2}.

The structure functions measured in DIS, determine only a combination
of quark distributions rather the individual quark distributions. Thus
it is possible, in principle, that models which give almost identical
results for the EMC ratio can give different predictions for the
Drell-Yan~\cite{chmaj84,chmaj85} process. The EMC effect has indeed been found
to be experimentally consistent (in the time-like region) with both
the pion- and proton-induced Drell-Yan
processes~\cite{na10,drell_yan,dyreview2,dyreview3}. Since the sea quark
distributions are less well known the results from proton-induced
Drell-Yan results are far less conclusive in verifying the EMC effect.

Pion-induced Drell-Yan processes are relatively more sensitive to the
EMC effect, specially the cross sections ratios
$\frac{\sigma^{DY}(\pi^-+A)}{\sigma^{DY}(\pi^-+\text{D})}$,
$\frac{\sigma^{DY}(\pi^-+A)}{\sigma^{DY}(\pi^-+\text{H})}$, and
$\frac{\sigma^{DY}(\pi^++A)}{\sigma^{DY}(\pi^-+A)}$, where $A$

represents a nuclear, D a deuteron and H a hydrogen target. Assuming
isospin symmetry, which implies $u_{\pi^+} = d_{\pi^-}$,
$\bar{u}_{\pi^-} = \bar{d}_{\pi^+}$, $\bar{u}_{\pi^+} =
\bar{d}_{\pi^-}$, $u_{\pi^-} = d_{\pi^+}$ and keeping only the
dominant terms in each cross--section, we get

\begin{equation}
\label{eq:RaD}
R^{-}_{A/D} = \frac{\sigma^{DY}(\pi^- + A)}{\sigma^{DY}(\pi^- + \text{D})}
\approx \frac{u_{A}(x)}{u_{D}(x)},
\end{equation}

\begin{equation}
\label{eq:RAH}
R^{-}_{A/H} = \frac{\sigma^{DY}(\pi^- + A)}{\sigma^{DY}(\pi^- + \text{H})}
\approx \frac{u_{A}(x)}{u_p(x)},
\end{equation}

\begin{equation}
\label{eq:Rpm}
R_{\pm} = \frac{\sigma^{DY}(\pi^+ + A)}{\sigma^{DY}(\pi^- + A)}
\approx \frac{d_{A}(x)}{4\,u_{A}(x)}.
\end{equation}

The target quark distributions have a subscript $A$, and the up quark
distribution in the deuteron and the proton are labeled by $u_{D}$ and
$u_p$, respectively. It is clear that pion-induced Drell-Yan
cross--section ratios, specially those between a heavy nuclear target
and a hydrogen or deuterium target, should be sensitive to the EMC
effect and, importantly, they are not sensitive to the pion structure
functions, which are not yet accurately determined.

\begin{figure}[hbt]
\centerline{\psfig{file=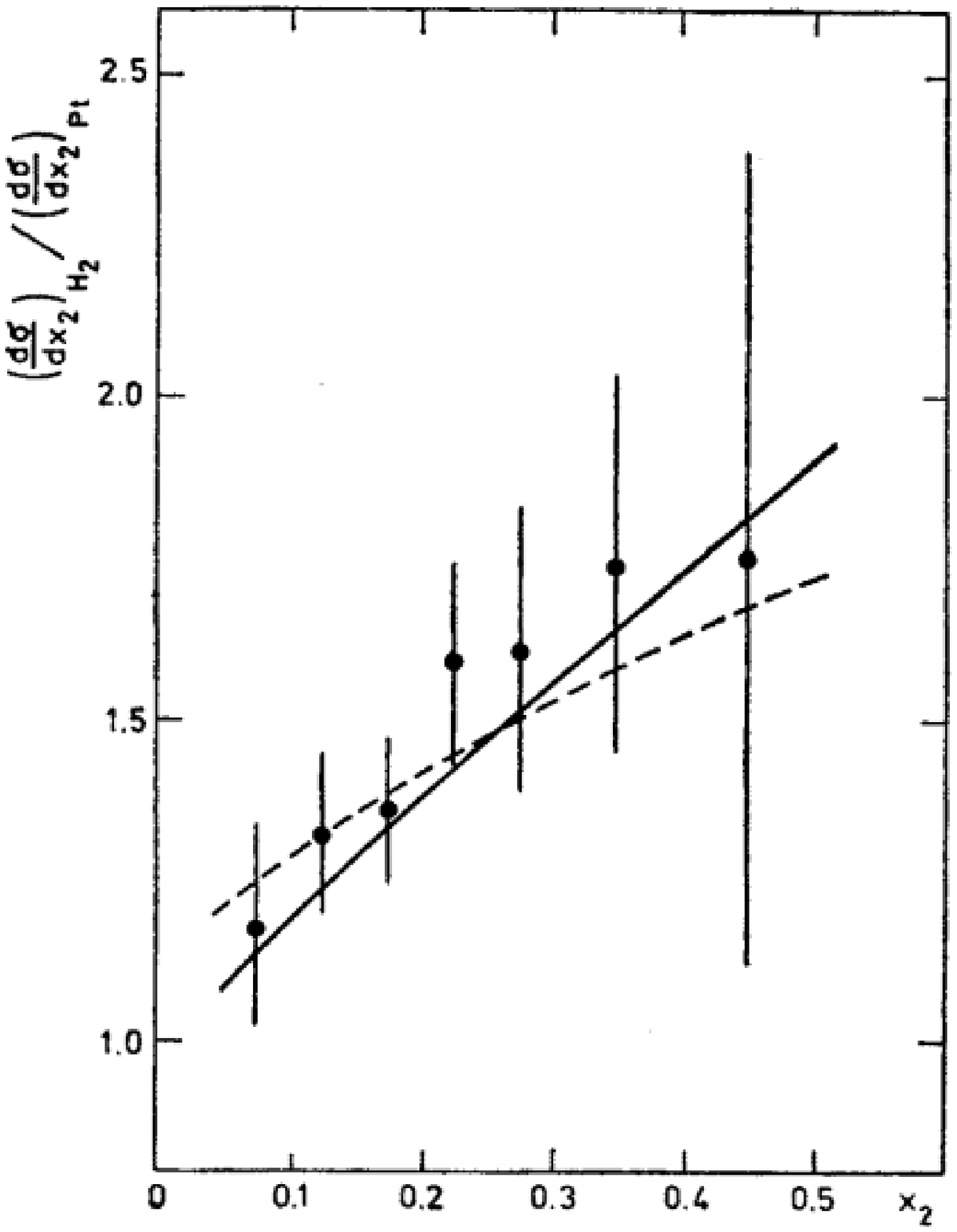,width=5.0cm}\psfig{file=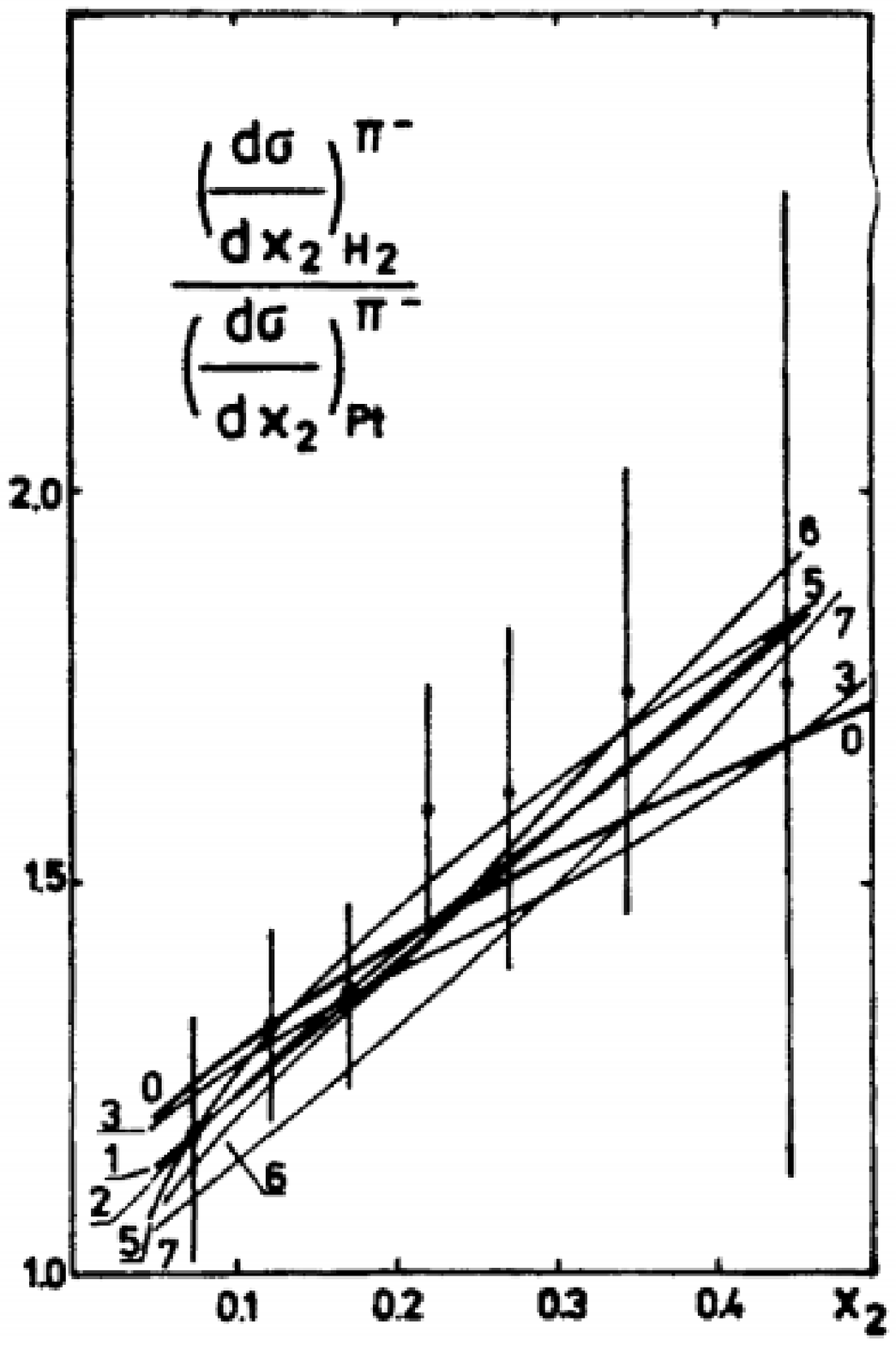,width=4.5cm}}
\caption{(left)The ratio
  $\frac{\sigma^{DY}(\pi^-+H)}{\sigma^{DY}(\pi^-+\text{Pt})}$ vs $x_2$
  the Bjorken scaling variable for the interacting quark of the
  nucleon in the target nucleus. The data are from the NA3 experiment
  and they are compared to a fit to the EMC data with (solid line) and
  without (dashed) corrections for $\Delta$ isobars. (right) The ratio
  $\frac{\sigma^{DY}(\pi^-+H)}{\sigma^{DY}(\pi^-+\text{Pt})}$ vs $x_2$
  from the NA3 data compared to seven different models which reproduce
  the EMC effect. Figures taken from Ref.~\protect \refcite{chmaj85}}
\label{na3_fig}
\end{figure}

The existing set of pion-induced Drell-Yan data that can be used to
form these ratios are primarily from three experiments; the NA3, the NA10 and the Omega experiments.
Chmaj and Heller~\cite{chmaj84,chmaj85} have compared the ratio
$\frac{\sigma^{DY}(\pi^-+H)}{\sigma^{DY}(\pi^-+\text{Pt})}$, measured
by the NA3 collaboration~\cite{na3}, with fits to the EMC data as well
as a wide variety of models that accurately reproduce the EMC
effect. They found that indeed the Drell-Yan cross section ratios are
sensitive to the EMC effect (see Fig.~\ref{na3_fig}), however, most
models available in 1985, which could reproduce the EMC data, gave
predictions for the Drell-Yan process which have little or no
difference. The results of their comparison of the NA3 data with a fit
to the EMC data and seven widely different models, that all reproduced
the EMC effect, is shown in Fig.~\ref{na3_fig}. They concluded that
the cross-section ratio obtained from the NA3 data confirm the
modification of the quark distributions inside the nucleus (i.e. the
EMC effect) however they are not precise enough to distinguish
between the various models of the EMC effect.

\begin{figure}[hbt]
\centerline{\psfig{file=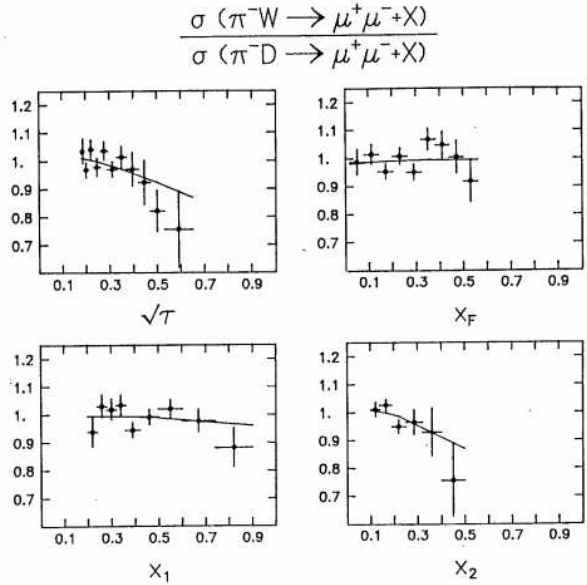,width=9.5cm}}
\caption{The ratio
  $\frac{\sigma^{DY}(\pi^-+W)}{\sigma^{DY}(\pi^-+\text{D})}$ for the
  combined $286\,$GeV and $140\,$GeV data~\cite{na10}, compared to the
  model predictions based on the BCDMS results~\cite{muons,muons87}.}
\label{na10_fig}
\end{figure}

The data from the NA10 experiment~\cite{na10} was used to form the
ratio $\frac{\sigma^{DY}(\pi^-+W)}{\sigma^{DY}(\pi^-+\text{D})}$ for
the $286\,$GeV and $140\,$GeV $\pi^-$ beam, as well as both data sets
combined together. In Fig.~\ref{na10_fig}, the ratio for the combined 
data set is shown vs $x_{1}$ the Bjorken scaling variable for the
interacting quark in the pion, $x_2$ the analogous quantity for the
nucleon in the target nucleus, $\sqrt{\tau} \approx \sqrt{x_1 x_2}$
and $x_F \approx x_1 - x_2$. The data are compared to the ratio of
structure functions obtained from the DIS experiment by the BCDMS
collaboration~\cite{muons,muons87}. The NA10 Drell-Yan data were completely
consistent with the modification of the nucleon quark distributions in
the nucleus, while the pion quark distributions were found to be
unaffected, all consistent with QCD factorization.

\section{Quark flavor--dependent EMC effect}

\label{sec:emcflavor}


The specific origins of the observed $A$ dependence of the nuclear
quark distributions have yet to be unambiguously identified. Attempts
to explain the EMC effect have led to an extensive collection of
theoretical models~\cite{geesaman_review,norton_review}, that describe
the essential features of the data, however the important physics of
these models is often very different. This suggests that there are
aspects of the EMC effect that are not probed in DIS, and new
experimental observables are essential to understand the origins of
the EMC effect. The quark flavor dependence of the EMC effect is one
such promising experimental observable. In the simplest picture where
the nuclear parton distribution functions (PDFs) are just the nucleon
PDFs smeared by the Fermi motion of the nucleons, and thus the nuclear
modification would be very similar for $u$ and $d$
quarks~\cite{ian}. In contrast, some models such as the pion-excess
model~\cite{pion1,pion2,pion3} predicts a flavor--dependent nuclear
modifications arising from the different isospin composition of the
pion cloud of protons and neutrons. Similarly, a recent model, the
Clo\"{e}t, Bentz and Thomas (CBT) reported in
Refs.~\refcite{Cloet:2006bq,ian}, claims that for $N\neq Z$ nuclei
(where $N$ and $Z$ refer to the number of neutrons and protons,
respectively) the isovector--vector mean field (usually denoted by
$\rho^0$) will affect the $u$ quarks differently from the $d$ quarks
in the bound nucleons. Therefore, this model predicts that the $u$ and
$d$ quarks have distinct nuclear modifications influenced by the $N/Z$
ratio of the nucleus~\cite{Cloet:2006bq}, leading to a flavor
sensitive EMC effect in nuclei with $N \neq Z$.

\begin{figure}[hbt]
\centerline{\psfig{file=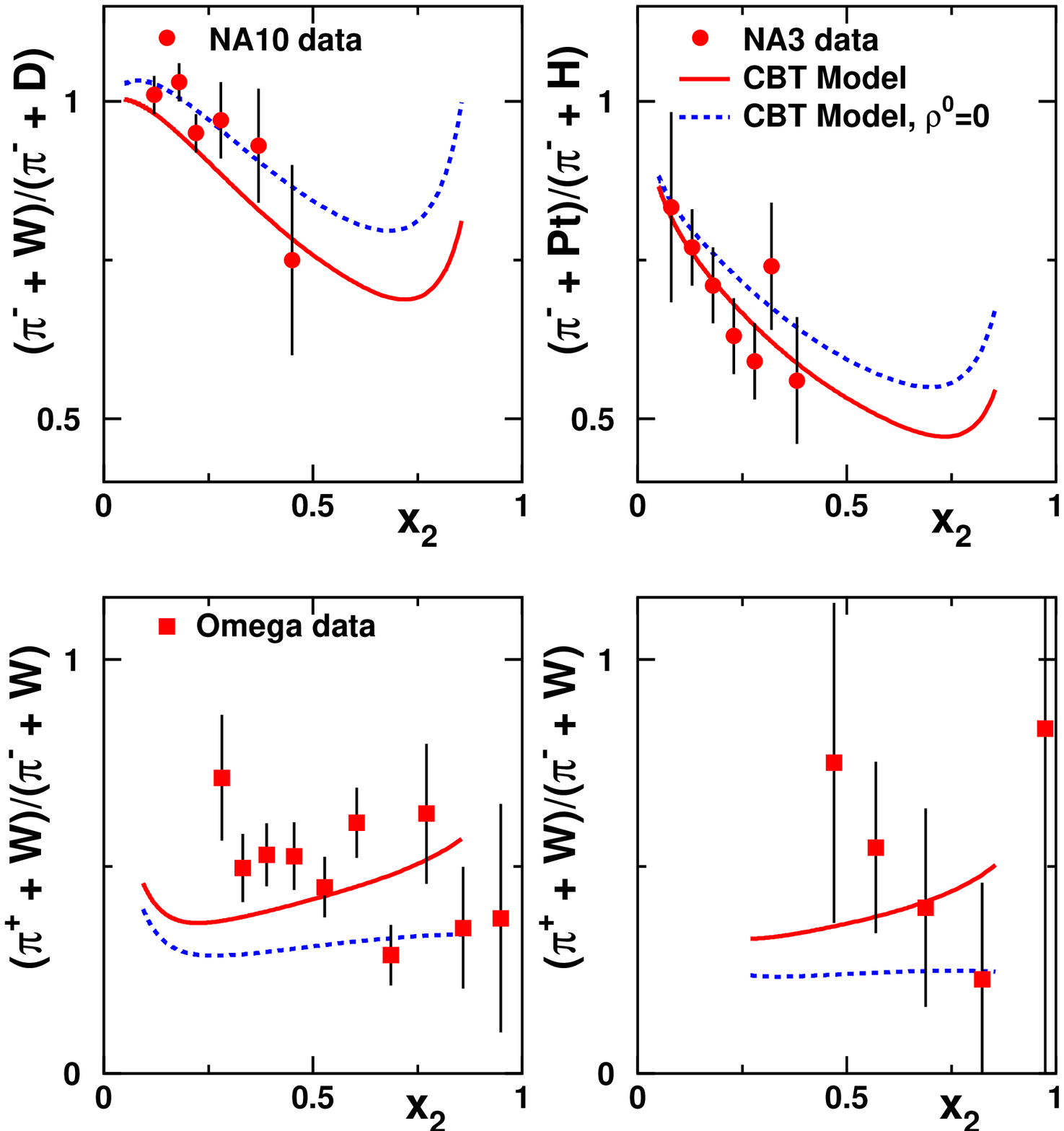,width=0.48\textwidth}\psfig{file=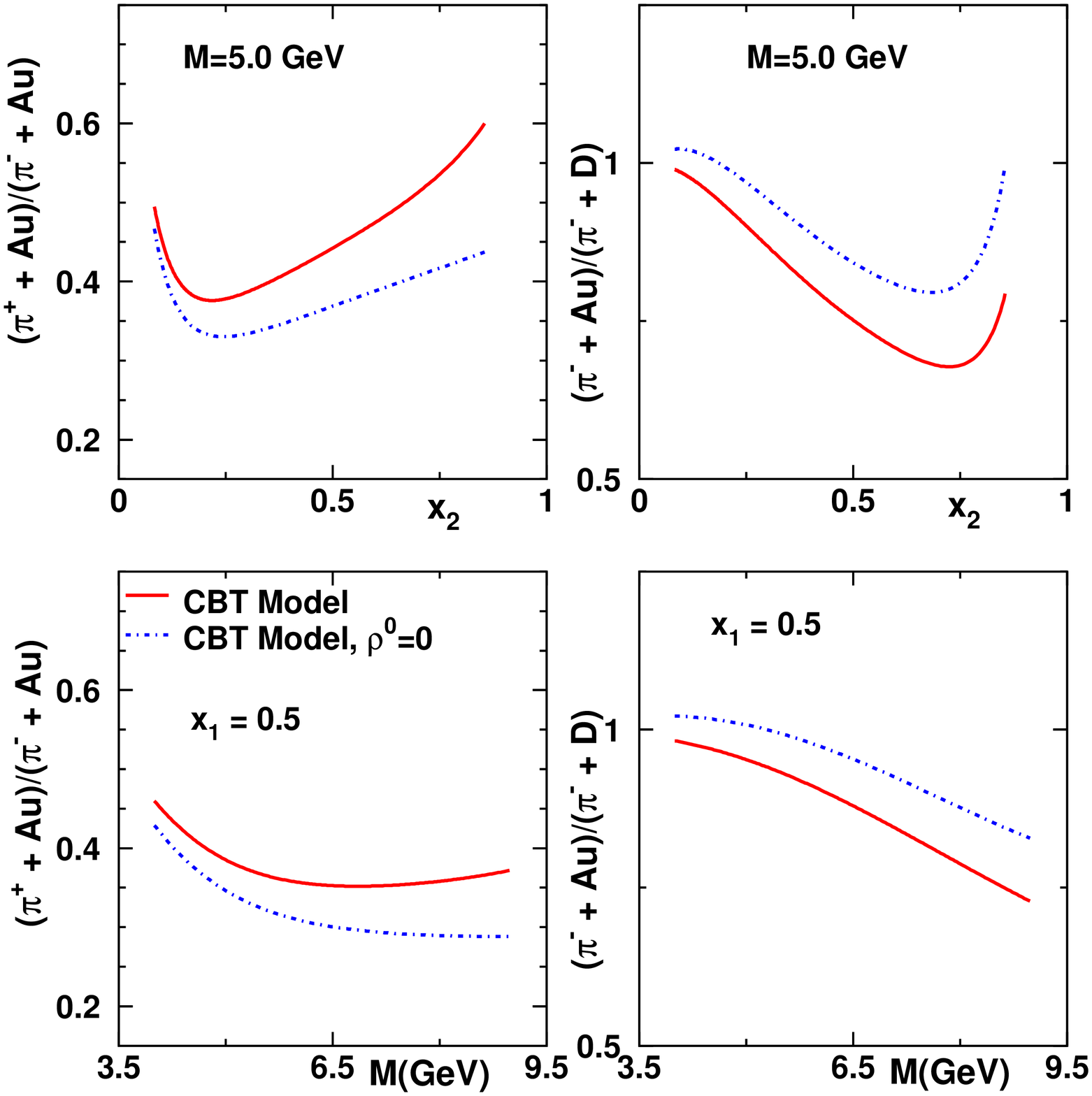,width=0.48\textwidth}}
\caption{The existing data for the ratios (a)
  $\frac{\sigma^{DY}(\pi^-+\text{W})}{\sigma^{DY}(\pi^-+\text{D})}$,
  (b) $\frac{\sigma^{DY}(\pi^-+\text{Pt})}
  {\sigma^{DY}(\pi^-+\text{H})}$ and
  $\frac{\sigma^{DY}(\pi^++\text{W})} {\sigma^{DY}(\pi^-+\text{W})}$
  for (c) $5.3 \leq Q^2 \leq 7.2\,$GeV$^2$ and (d) $16 \leq Q^2 \leq
  25\,$GeV$^2$. The curves are results using the nuclear PDFs from the
  CBT model with $N/Z$ equal to that of tungsten. The solid line is
  the full flavor--dependent result and the dashed line is obtained by
  setting the $\rho^{0}$ mean--field to zero. The Drell-Yan
  cross--section ratios for (e)
  $\frac{\sigma^{DY}(\pi^++\text{Au})}{\sigma^{DY}(\pi^-+\text{Au})}$
  and (f)
  $\frac{\sigma^{DY}(\pi^-+\text{Au})}{\sigma^{DY}(\pi^-+\text{D})}$
  as a function of $x_2$, using nuclear PDFs from the CBT model for
  gold (red solid) and the CBT model with the $\rho^{0}$ mean--field
  set to zero (blue dot-dashed). (g) and (h) show the same ratios as a
  function of the dimuon mass, at a fixed $x_1 = 0.5$. Figures taken
  from Ref.~\protect \refcite{dutta11}.}
\label{fig_dutta2}
\end{figure}

Pion-induced Drell-Yan processes are an experimental tool that is
sensitive to flavor--dependent effects in the nuclear quark
distributions. We have already noted that the cross section ratios of
pionic Drell-Yan processes (Eqs.~\eqref{eq:RaD}--\eqref{eq:Rpm}) are
sensitive to the EMC effect. Moreover, they are also sensitive to the
flavor dependence of the EMC effect. The Drell-Yan ratio, $R_\pm$, of
Eq.~\eqref{eq:Rpm} is an ideal experimental observable to search for
flavor--dependent EMC effect, since it is directly proportional to
$d_A(x)/u_A(x)$.  The only existing data on $R_\pm$ are from the Omega
collaboration~\cite{omega}. In Ref.~\refcite{dutta11}, the CBT model
with and without the flavor dependence was compared to the existing
pionic Drell-Yan data as shown in Fig.~\ref{fig_dutta2} (a--d). From the
analysis of Ref.~\refcite{dutta11} it is clear that the
flavor--dependent model is preferred for the NA3 and the Omega
collaborations data sets, whereas the NA10 results are better
described by the flavor--independent CBT model.  Unfortunately, the
existing data lack the precision needed to place a useful constraint
on the flavor dependence of the EMC effect.

The COMPASS collaboration~\cite{compass} proposes to measure the
Drell-Yan cross--sections with 190-GeV$/c$ pion beams which will be an
opportunity to test the flavor--dependence of the EMC effect. The
sensitivity of the proposed COMPASS measurement was examined in
Ref.~\refcite{dutta11} and is shown in
Fig.~\ref{fig_dutta2}(e--h). The significant difference between the
predicted ratios for the flavor--dependent versus flavor--independent
nuclear PDFs provide a strong motivation for these future
measurements.


\section{Future Programs}

\label{sec:future}


\subsection{Polarized pion-induced Drell-Yan in COMPASS-II at CERN}

As mentioned in Sec.~\ref{sec:angdist2_theory} the angular
distributions of leptons from Drell-Yan process may be sensitive to a
key component of the TMD parton distributions known as the
Boer-Mulders function~\cite{BM}. The other important single spin
asymmetry (SSA) component in TMD parton distributions which could also
be explored by the Drell-Yan process is the Sivers function
($f_{1T}^{\perp}$ )~\cite{Sivers}. This function describes how the
transverse momentum $k_T$ distribution of unpolarized partons is
distorted by the transverse polarization of the parent hadron. The
correlation between $k_T$ and parton/hadron transverse polarization is
intuitively resulted from a non-vanishing orbital angular momentum of
the quarks themselves.

Sivers function, formerly considered to be zero by the time-reversal
property of QCD~\cite{Collins:1992}, was found to survive because of
the presence of a gauge line (Wilson line) for the need of gauge
invariance in the definition of TMD parton
distribution~\cite{Collins}. This single spin asymmetry has been
clearly observed in the semi-inclusive deeply inelastic scattering
process (SIDIS) by HERMES experiment at DESY~\cite{HERMES:SIDIS,hermes2,hermes3} and
COMPASS experiment at CERN~\cite{COMPASS:SIDIS,compass2,compass3,compass4}. That has allowed an
extraction of Sivers function with global fits~\cite{Sivers:PDF,sivers2,sivers3,sivers4} and
opposite contributions from $u$ and $d$ quarks are seen.

Sivers function could also be extracted from the polarized Drell-Yan
process with the great advantage of it being free from convolution of
fragmentation function. There exists an essential prediction that the
$f_{1T}^{\perp}$ functions extracted from Drell-Yan processes and
those obtained from SIDIS must be reserved in sign because the Wilson
line happens in the initial and final state
respectively~\cite{Collins}. The same prediction also applies for the
Boer-Mulders function to be determined in SIDIS and Drell-Yan
processes. An experimental verification of the sign-reversal property
of the Sivers and Boer-Mulders functions is a crucial test of QCD in
the non-perturbative regime on the aspects of the origin of SSAs and
the validity of factorization scheme -- the corner stone of TMD
physics.

A proposal~\cite{compass} to carry out a polarized Drell-Yan
measurement by COMPASS experiment in 2014 at CERN was approved. The
angular distributions of Drell-Yan events produced by 190-GeV/$c$
$\pi^-$ beam colliding with transversely polarized NH$_3$ target will
be measured~\cite{COMPASS:DY,compass:dy2}. Both the Sivers and Boer-Mulders
functions from the target quark and the beam anti-quark could be
extracted from the azimuthal spin asymmetries for testing the
universality of the Sivers and Boer-Mulders functions between
Drell-Yan and SIDIS. It should be noted that an unpolarized Drell-Yan
program using liquid hydrogen, liquid deuterium and nuclei targets in
COMPASS experiment will be very helpful to further investigate the
known interesting phenomena in the dilepton angular distributions and
the EMC effect.

\subsection{Exclusive pion-induced Drell-Yan at J-PARC}

\begin{figure}[hbt]
\centerline{\psfig{file=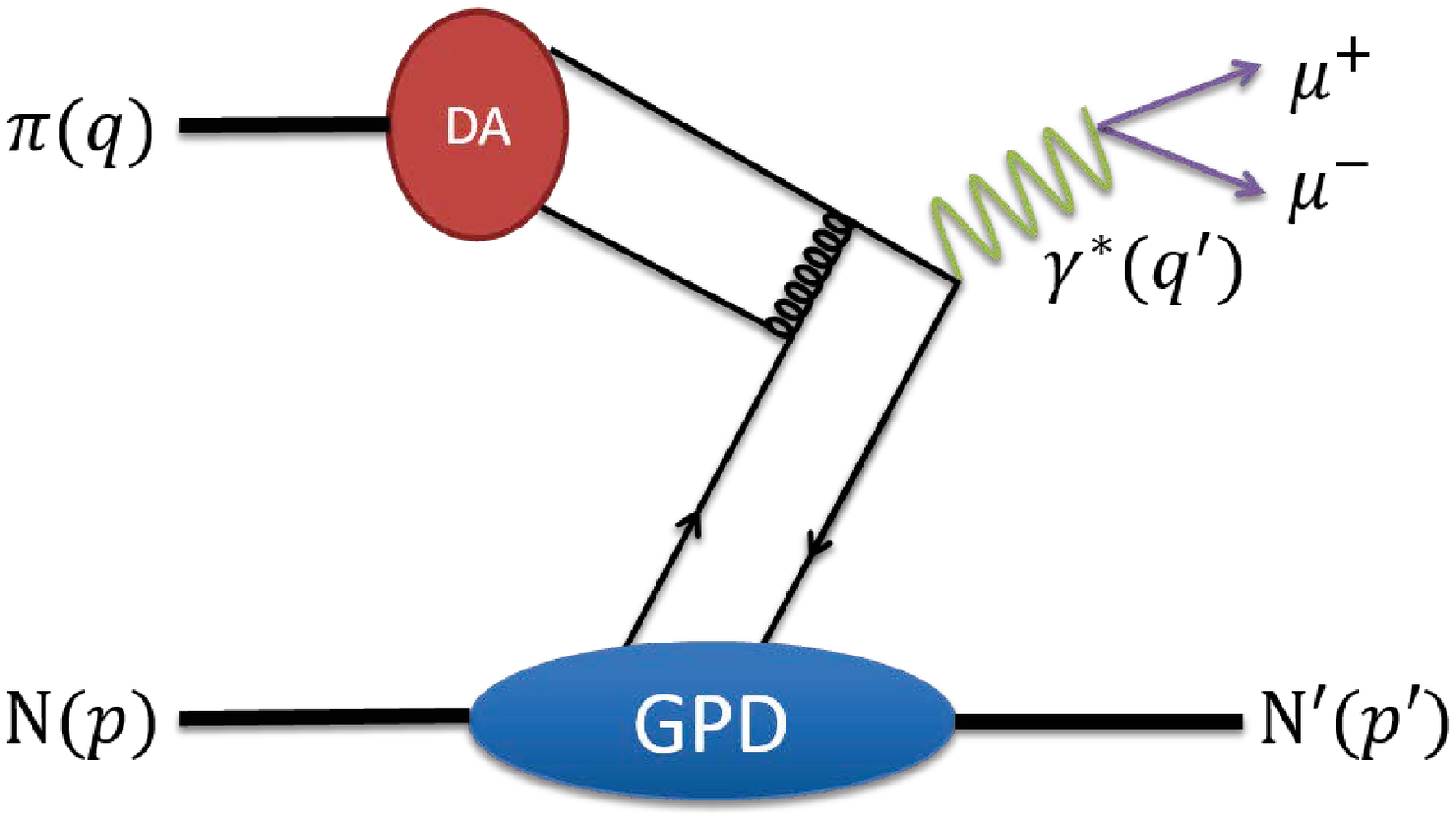,width=0.45\textwidth}\psfig{file=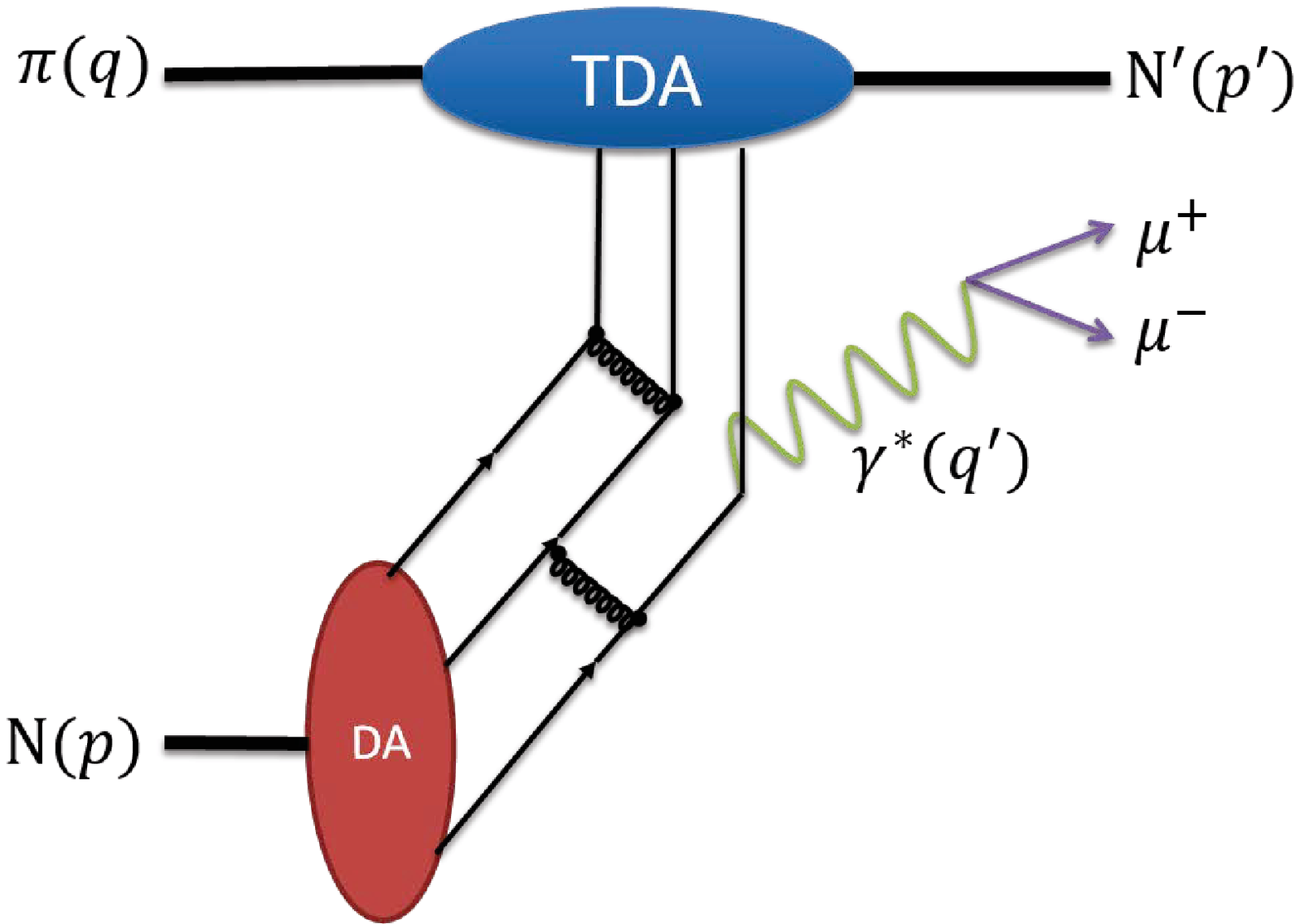,width=0.45\textwidth}}
\caption{Exclusive pion-induced $2 \rightarrow 2$ processes: with (a)
  small and (b) large momentum-transfer to the nucleon
  target~\cite{Pire:pionDA,pire12,pire05,Pire:pionTDA,pire07,pire2:07,pire08,pire3:07,pire4:07}.}
\label{fig:pionGDA}
\end{figure}

When $x_{\pi}$ approaches the limit of $1$, the inclusive Drell-Yan
process actually becomes an exclusive one: $\pi^- N \rightarrow N'
\mu^+ \mu^-$. There are recent theoretical studies on the exclusive
pion-induced Drell-Yan process where the nucleon remains intact in the
final state~\cite{Pire:pionDA,pire12,pire05,Pire:pionTDA,pire07,pire2:07,pire08,pire3:07,pire4:07}. This process is a
time-like crossing of the deeply virtual pion production process. A
factorization of this process shows that three important
non-perturbative components are involved: nucleon GPD, pion DA and
nucleon-to-pion transition distribution amplitude (TDA), as shown in
Fig.~\ref{fig:pionGDA}.

The GPDs contain information of the correlation between the
quark/gluon transverse position in the nucleon and its longitudinal
momentum, which has been assessed by deeply virtual Compton scattering
(DVCS) and deep virtual meson production (DVMP). The distribution
amplitudes of nucleons probe the three-quark component of the nucleon
light-cone wave function, while higher order components in the
Fock-space expansion of the nucleon state are essential to describe
the nucleon-to-pion transition distribution amplitudes (TDA)
~\cite{Pasquini}. When studying the nucleon structure, the
nucleon-to-pion TDAs are particularly interesting because they
directly probe the three-quark plus sea $q\bar{q}$ pair component,
$\psi_{(3q+q\bar{q})}$, which is related to the pion-cloud component
inside the nucleon.

Since the cross sections of exclusive processes is larger at low CMS
energies, it is appropriate to carry out such kind of measurement with
pion beam of lower momentum. An approved three-year project of
constructing a high-momentum beam line (HiPBL) in the Hadron Hall at
J-PARC~\cite{JPARC} started from year 2013. Upon completion of HiPBL,
high-flux primary proton beam and secondary particles, e.g.  $\pi$,
$K$ or $\bar{p}$ in the momentum range of $15-20\,$GeV/$c$ would be
available for the study of hadron physics. An exclusive Drell-Yan
process could be experimentally characterized if the scattered
nucleons in the final state be either directly detected, or identified
via the missing-mass technique. This measurement is very challenging
considering the very small production cross section in the order of
pico barn and a requirement of good momentum resolution for the
decayed leptons passing through hadron absorber. Nevertheless the
results in the time-like region, if available, will be complementary
to those in space-like region assessed by the DVCS and DVMS processes
to be studied in JLab~\cite{JLAB}.


\section{Conclusion}

\label{sec:conclusion}


We summarize the results of dilepton angular distributions and the
nuclear modification of the parton distributions from the pion-induced
Drell-Yan process. The discrepancies between the observed dilepton angular distributions and 
the expectations of the Drell-Yan parton model together with perturbative QCD corrections, 
can be resolved by incorporating the non-perturbative QCD physics associated with the 
bound-state effects of the partons.
These connections open up a clean and important way of accessing
the TMD distribution and GPD of nucleons, and pion DA. Also the study
of pion-induced Drell-Yan process on heavy nuclei will help reveal
the very interesting flavor-dependent modification of quark
distributions in the nuclear medium. With several upcoming experiments
and others being planned, we expect that pion-induced Drell-Yan
process should continue improving our understanding of QCD and the
hadron structure in multi-dimensions.

\section*{Acknowledgments}

This work is supported in part by the U.S. Department of Energy (Grant
No. DE-FG02-07ER41528) and the National Science Council of Taiwan
(Grant No. 100-2112-M-001-015-MY3).



\begin{thebibliography}{0}



\bibitem{christenson70} J. Christenson {\it et al.},
  Phys. Rev. Lett. {\bf 25}, 1523 (1970).
\bibitem{christenson73} J. Christenson {\it et al.},
  Phys. Rev. {\bf D8}, 2016 (1973);



\bibitem{drell-yan1} S.D.~Drell and T.M.~Yan Phys. Rev. Lett. {\bf
  25}, 316 (1970).
\bibitem{drell-yan2}S.D.~Drell and T.M.~Yan Phys. Rev. Lett. {\bf
  25}, 902 (1970).



\bibitem{dyreview1} I.R.~Kenyon, Rep. Prog. Phys. {\bf 45}, 1261
  (1982).
\bibitem{dyreview2}C.~Grosso-Pilcher and M.J.~Shochet,
  Annu. Rev. Nucl. Part. Sci. {\bf 36}, 1 (1986). 
\bibitem{dyreview3}P.L.~McGaughey,
  J.M.~Moss, and J.C.~Peng, Annu. Rev. Nucl. Part. Sci. {\bf 49},
  217 (1999).



\bibitem{PDF} E.~Perez and E.~Rizvi, Rep. Prog. Phys. {\bf 76}, 046201
  (2013) and references therein.



\bibitem{TMD} V.~Barone, F.~Bradamante, A.~Martin,
  Prog. Part. Nucl. Phys. {\bf 65}, 267 (2010) and references therein.



\bibitem{HEPdata} Durham database of pionic Drell-Yan production,
  \url{http://durpdg.dur.ac.uk/spires/hepdata/online/dy/pi-Nto2muX.html}
  and
  \url{http://durpdg.dur.ac.uk/spires/hepdata/online/dy/pi+Nto2muX.html}.



\bibitem{CIP1} C.B.~Newman {\it et al.} (CIP Collaboration),
  Phys. Rev. Lett. {\bf 42},948 (1979).



\bibitem{CIP2} K.J.~Anderson {\it et al.} (CIP Collaboration),
  Phys. Rev. Lett. {\bf 43},1219 (1979). 
\bibitem{CIP3}S.~Palestini {\it et al.}
  (CIP Collaboration), Phys. Rev. Lett. {\bf 55},2649 (1985).
\bibitem{CIP4}J.G.~Heinrich {\it et al.} (CIP Collaboration), Phys. Rev. D{\bf
    44},1909 (1991).



\bibitem{NA3} J.~Badier {\it et al.} (NA3 Collaboration), Z. Phys. C
  {\bf 11}, 195 (1981).



\bibitem{na3}A. Michelini, CERN-EP/81-128 (1981); J. Badier {\it et
  al.} (NA3 Collaboration), Phys. Lett. B {\bf 104}, 335 (1981).

%

\bibitem{omega}M. Corden {\it et al.} (Omega Collaboration),
  Phys. Lett. B {\bf 96}, 417 (1980).



\bibitem{NA10:86} S.~Falciano {\it et al.} (NA10 Collaboration),
  Z. Phys. C {\bf 31}, 513 (1986). 
\bibitem{NA10:88}M.~Guanziroli {\it et al.} (NA10
  Collaboration), Z. Phys. C {\bf 37}, 545 (1988).



\bibitem{na10} P.~Bordalo {\it et al.} (NA10 Collaboration), Phys. Lett. B {\bf 193}, 368 (1987).



%

\bibitem{E615} J.S.~Conway {\it et al.} (E615 Collaboration),
  Phys. Rev. D {\bf 39}, 92 (1989).



\bibitem{GPD} M.~Diehl, Phys. Rept. {\bf 388}, 41 (2003) and
  references therein.



\bibitem{LamTung78} C.S.~Lam and W.K.~Tung, Phys.\ Rev.\ D {\bf 18},
  2447 (1978). 
\bibitem{LamTung79}C.S.~Lam and W.K.~Tung, Phys.\ Lett.\ B {\bf 80}, 228
  (1979).
\bibitem{LamTung80}C.S.~Lam and W.K.~Tung, Phys.\ Rev.\ D {\bf 21}, 2712
  (1980).



\bibitem{Nachtmann} A.~Brandenburg, O.~Nachtmann, and E.~Mirkes,
  Z. Phys. C {\bf 60}, 697 (1993).



\bibitem{LT_NNLO} E.~Mirkes and J.~Ohnemus, Phys. Rev. D {\bf 51}, 4891(1995)



\bibitem{Berger} E.L.~Berger and S.J.~Brodsky, Phys. Rev. Lett. {\bf
  42}, 940 (1979).



\bibitem{Brandenburg} A.~Brandenburg, S.J.~Brodsky, V.V.~Khoze, and
  D.~M\"{u}ller, Phys.\ Rev.\ Lett {\bf 73}, 939 (1994).



\bibitem{TFF} X.-G.~Wu, T.~Huang, and T.~Zhong, Chin. Phys. C {\bf
  37}, 063105 (2013)



\bibitem{Bakulev} A.P.~Bakulev N.G.~Stefanis, and O.V.~Teryaev,
  Phys.\ Rev.\ D {\bf 76}, 074032 (2007)



\bibitem{Boer05} D.~Boer, A.~Brandenburg, O.~Nachtmann, and
  A.~Utermann, Eur. Phys. J. {\bf C40}, 55 (2005).



\bibitem{Nachtmann2} O.~Nachtmann and A.~Reiter, Z. Phys. C {\bf 24},
  283 (1984).
\bibitem{Nachtmann3} G.W.~Botz, P.~Haberl, and O.~Nachtmann, Z. Phys. C {\bf
    67}, 143 (1995).



\bibitem{Boer:LT} D.~Boer, Phys.\ Rev.\ D {\bf 60}, 014012 (1999).



\bibitem{BM} D.~Boer and P.J.~Mulders, Phys.\ Rev.\ D {\bf 57}, 5780
  (1998).



\bibitem{E866:p} L.Y.~Zhu {\it et al.} (FNAL E866/NuSea Collaboration),
  Phys.\ Rev.\ Lett. {\bf 99}, 082301 (2007).



\bibitem{E866:d} L.Y.~Zhu {\it et al.} (FNAL E866/NuSea Collaboration),
  Phys.\ Rev.\ Lett.\ {\bf 102}, 182001 (2009).



\bibitem{CDF} T. Aaltonen {\it et al.} (CDF Collaboration),
  Phys. Rev. Lett. {\bf 106}, 241801 (2011)



\bibitem{Instanton} A.~Brandenburg, A.~Ringwald, and
  A.~Utermann, Nucl. Phys. B {\bf 754}, 107 (2005).



\bibitem{GlauberGluon} C.P.~Chang and H.N.~Li, arXiv:1305.4694



\bibitem{emc1} J.J.~Aubert {\it et al.}  (European Muon
  Collaboration), Phys.\ Lett.\ B {\bf 123}, 275 (1983).



\bibitem{slac} J.~Gomez {\it et al.}, Phys.\ Rev.\ D {\bf 49}, 4348
  (1994).



\bibitem{muons}G. Bari {\it et al.}, Phys. Lett. B {\bf 163}, 282
  (1985).
\bibitem{muons87}A.C.~Benvenuti {\it et al.}, Phys. Lett. B {\bf 189}, 483
  (1987).



\bibitem{neutrinos}A.M.~ Cooper {\it et al.}, Phys. Lett. B {\bf
  141}, 133 (1984).
\bibitem{neutrinos2} H.~Abramowicz {\it et al.}, Z. Phys. C {\bf 25},
  29 (1984).



\bibitem{hermes} K.~Ackerstaff {\it et al.}  (HERMES Collaboration),
  Phys.\ Lett.\ B {\bf 475}, 386 (2000). 



\bibitem{seely09}J. Seely {\it et al.}, Phys. Rev. Lett. {\bf 103},
  202301 (2009). 






\bibitem{chmaj84} T.~Chmaj and K.J.~Heller, Acta Phy. Polo. {\bf
  B15}, 473 (1984).
\bibitem{chmaj85} T.~Chmaj and K.J.~Heller, Acta Phy. Polo. {\bf
  B16}, 423 (1985).



\bibitem{drell_yan}D.M.~Alde {\it et al.} (E772 Collaboration), Phys. Rev. Lett. {\bf 64}, 
2479 (1990).



\bibitem{geesaman_review} D.F.~Geesaman, K.~Saito and A.W.~Thomas,
  Ann.\ Rev.\ Nucl.\ Part.\ Sci.\ {\bf 45}, 337 (1995).



\bibitem{norton_review} P.R.~Norton, Rept.\ Prog.\ Phys.\ {\bf 66},
  1253 (2003).



\bibitem{ian}I.C.~Clo\"{e}t, W.~Bentz and A.W.~Thomas,
  Phys. Rev. Lett. {\bf 102}, 252301 (2009).




\bibitem{pion1} C.H.~Llewellyn-Smith, Phys. Lett. B {\bf 128}, 107
  (1983).

%

\bibitem{pion2} M.~Ericson and A.W.~Thomas, Phys. Lett. B {\bf 128},
  112 (1983).

%

\bibitem{pion3} E.L.~Berger, F.~Coester, and R.B.~Wiringa,
  Phys. Rev. D {\bf 29}, 398 (1984).



\bibitem{Cloet:2006bq}

  I.C.~Cloet, W.~Bentz, A.W.~Thomas,
  Phys.\ Lett.\  {\bf B642}, 210-217 (2006).



\bibitem{dutta11} D.~Dutta, J.-C.~Peng, I.C.~Cloet and D.~Gaskell,
  Phys. Rev. C {\bf 83}, 042201R (2011).



\bibitem{compass} COMPASS Collaboration, CERN Report No. CERN/SP SLC
  96-14, SPSC/P 297; The COMPASS-II Proposal, \url{http://wwwcompass.cern.ch/compass/proposal/compass-II_proposal/compass-II_proposal.pdf}



\bibitem{Sivers} D.~Sivers, Phys.\ Rev.\ D {\bf 41}, 83 (1990);
  D.~Sivers, Phys.\ Rev.\ D {\bf 43}, 261 (1991).



\bibitem{Collins:1992} J.C.~Collins, Nucl. Phys. {\bf B 396}, 161 (1993).



\bibitem{Collins} J.C.~Collins, Phys.\ Lett.\ B {\bf 536}, 43 (2002).



\bibitem{HERMES:SIDIS} A.~Airapetian {\it et al.} (HERMES
  Collaboration), Phys. Rev. Lett. {\bf 94}, 012002 (2005).
\bibitem{hermes2}L.L.~Pappalardo et al. (HERMES Collaboration), Eur. Phys. J. A {\bf
    38}, 145 (2008).
\bibitem{hermes3} A.~Airapetian {\it et al.} (HERMES
  Collaboration), Phys. Rev. Lett. {\bf 103}, 152002 (2009).



\bibitem{COMPASS:SIDIS} V.Y.~Alexakhin, {\it et al.} (COMPASS
  Collaboration), Phys. Rev. Lett. {\bf 94}, 202002 (2005).
\bibitem{compass2}M.~Alekseev {\it et al.} (COMPASS Collaboration), Phys. Lett. B {\bf 673},
  127 (2009).
\bibitem{compass3}M.G.~Alekseev {\it et al.} (COMPASS Collaboration),
  Phys. Lett. B {\bf 692}, 240 (2010).
\bibitem{compass4} C.~Adolph {\it et al.} (COMPASS
  Collaboration), Phys. Lett. B {\bf 717}, 383 (2012).



\bibitem{Sivers:PDF} W.~Vogelsang and F.~Yuan, Phys. Rev. D {\bf 72},
  054028 (2005).
\bibitem{sivers2}J.C.~Collins {\it et al.}, Phys. Rev. D {\bf 73},
  014021 (2006). 
\bibitem{sivers3}M.~Anselmino {\it et al.}, Eur. Phys. J. A {\bf 39},
  89 (2009).
\bibitem{sivers4}M.~Anselmino {\it et al.}, Phys. Rev. D {\bf 86}, 014028
  (2012).




\bibitem{COMPASS:DY} O.~Denisov (COMPASS Collaboration),
  Mod. Phys. Lett. A {\bf 24}, 3033 (2009).
\bibitem{compass:dy2}C.~Quintans (COMPASS
  Collaboration), J. Phys. Conf. Ser. {\bf 295}, 012163 (2011).



\bibitem{Pire:pionDA} E.R.~Berger, M.~Diehl, and B.~Pire,
  Phys. Lett. B {\bf 523}, 265 (2001). 
\bibitem{pire12}D.~Muller, B.~Pire,
  L.~Szymanowski, and J.~Wagner, Phys. Revs. D {\bf 86}, 031502
  (2012). 
\bibitem{pire05}B.~Pire and L.~Szymanowski, Phys. Lett. B 622 (2005) 83.



\bibitem{Pire:pionTDA} B. Pire and L. Szymanowski, Phys. Lett. B 622,
  83 (2005).
\bibitem{pire07}J.P.~Lansberg, B.~Pire, and L.~Szymanowski, Phys. Rev. D
  76, 111502(R) (2007).
\bibitem{pire2:07}J.P.~Lansberg, B.~Pire, and L.~Szymanowski,
  Phys. Rev. {\bf D 75}, 074004 (2007).
\bibitem{pire08}J.P.~Lansberg, B.~Pire, and L.~Szymanowski, Phys. Rev. {\bf 77}, 019902(E) (2008).
\bibitem{pire3:07}J.P.~Lansberg, B.~Pire, and L.~Szymanowski, AIP Conf. Proc. 892, 278 (2007).
\bibitem{pire4:07}J.P.~Lansberg, B.~Pire, and L.~Szymanowski, arXiv:0709.2567.



\bibitem{Pasquini}B.~Pasquini, M.~Pincetti, and S.~Boffi,
  Phys. Rev. {\bf D 80}, 014017 (2009).



\bibitem{JPARC} S.~Sawada, talk of ``Hadron experimental hall and
  high-momentum beamline at J-PARC'' in KEK theory center workshop on
  Hadron physics with high-momentum hadron beams at J-PARC in 2013.



\bibitem{JLAB} J.~Dudek {\it et al.}, Eur. Phys. J. A {\bf 48}, 187 (2012).



\end{thebibliography}
\end{document}